\documentclass[aps,prd,superscriptaddress,12pt,notitlepage]{revtex4}
\usepackage{epsfig}
\usepackage{graphicx}
\usepackage[font=small,skip=0pt,justification=raggedright]{caption}
\usepackage{amsmath,amssymb,mathrsfs}
\newcounter{fig}   \newcommand{\lbfig}[1]{\refstepcounter{fig}
\label{#1} }

\newcommand{\Tr}{{\rm Tr}}

\newcommand{\bea}{\begin{eqnarray}}
\newcommand{\eea}{\end{eqnarray}}
\newcommand{\be}{\begin{equation}}
\newcommand{\ee}{\end{equation}}

\newcommand{\re}[1]{(\ref{#1})}

\newcommand{\tr}{\mbox{tr}}

\newcommand{\eqn}{\begin{eqnarray}}
\newcommand{\eqnx}{\end{eqnarray}}

\date{\today}

\begin{document}
\title{Black holes with Skyrmion-anti-Skyrmion hairs}
\author{Ya. Shnir}
\affiliation{BLTP, JINR, Joliot-Curie 6, Dubna 141980, Moscow Region, Russia}
\begin{abstract}
We construct static axially symmetric black holes in multi-Skyrmion configurations
coupled to Einstein gravity in four dimensional asymptotically flat space-time.
In a simplest case the event horizon is located in-between a
Skyrmion-anti-Skyrmion pair, other solutions represent  black holes with
gravitationally bounded chains of Skyrmions and anti-Skyrmions placed along the axis
of symmetry in alternating order. We discuss the properties of these hairy
black holes and exhibit their domain of existence.

\end{abstract}
\maketitle

\section{Introduction}
Various black holes  with scalar hair, which
circumvent the well-known "no-hair" theorem (see, e.g.,
\cite{Herdeiro:2015waa,Volkov:2016ehx} and references therein),
are rather a common presence in the
landscape of gravity solutions. Historically, one of the first counter-examples
to the "no-hair" theorem was found in the Skyrme model
coupled to the Einstein gravity \cite{Luckock:1986tr,Droz:1991cx,Bizon:1992gb}.
It was shown that a small Schwarzschild black hole can be continuously connected
to the self-gravitating Skyrmion with some amount of the scalar field
absorbed into horizon. Another
example of the primary hairy black holes was obtained in the $SU(2)$
Einstein-Yang-Mills (EYM) theory \cite{Volkov:1989fi,Volkov:1990sva} and in the
Einstein-Yang-Mills-Higgs (EYMH) model \cite{Lee:1991vy,Breitenlohner:1994di}.
Interestingly, these static spherically symmetric solutions
share many common features with the corresponding black holes with Skyrme hairs, see e.g.
\cite{Volkov:2016ehx,Volkov:1998cc}.

There are various extensions of this type of the solutions. First, there are static axially
symmetric black holes both in the Einstein-Skyrme theory \cite{Sawado:2004yq} and in the EYM model
\cite{Kleihaus:1997ic}. Secondly, there are stationary spinning asymptotically flat hairy black holes in the
Einstein-Skyrme model \cite{Herdeiro:2018daq} and in the non-linear $O(3)$ sigma model \cite{Herdeiro:2018djx}
which belong to the same class of solutions, as the spinning black holes with primary scalar hair in the
Einstein-Klein-Gordon theory \cite{Hod:2012px,Herdeiro:2014goa}. Further, both the Skyrme theory
\cite{Krusch:2004uf,Shnir:2009ct,Shnir:2015aba}
and the EYMH model \cite{Kleihaus:1999sx,Kleihaus:2000hx,Kleihaus:2003nj,Kleihaus:2003xz,Kleihaus:2004is,Teh:2004bq,Paturyan:2004ps,Kleihaus:2004fh,Kleihaus:2005fs,Kunz:2006ex,Kunz:2007jw}
 admit axially symmetric equilibrium configurations with a
number of constituents located symmetrically with respect to the origin along the symmetry axis.
An example of such a configuration in the flat space is a sphaleron solutions that represent
a monopole-antimonopole pair in a
static equilibrium \cite{Kleihaus:1999sx,Kleihaus:2000hx}. Similar Skyrmion-anti-Skyrmion solution exists in the
Skyrme model \cite{Krusch:2004uf}. Further, there are generalizations of these solutions that represent
chains of interpolating Skyrmion-anti-Skyrmion \cite{Shnir:2009ct} and
monopole-antimonopole chains \cite{Kleihaus:2003nj,Kleihaus:2003xz,Kleihaus:2004is}.
Notably, inclusion of gravity allows for an attractive phase of Skyrmions not present in flat space
\cite{Shnir:2015aba}, flat space Skyrmion-anti-Skyrmion chains do not exists if the
topological charge of a constituent is less than two \cite{Krusch:2004uf,Shnir:2009ct}.

The similarity between the regular gravitating soliton configurations
of the non-abelian YMH theory and the Skyrme models
can also be noted in the pattern of their evolution. In both cases, there are
two branches of solutions, one of which emerges smoothly from the corresponding flat space configuration.
This branch is extended
up to some critical value of the effective gravitational coupling at which it merges the second, backward branch
leading to some limiting rescaled Bartnik-McKinnon type solutions \cite{Bartnik:1988am}.
Heuristically, this property can be attributed to the structure of the emerging
effective gravitational constant, which may
approach zero in two different limits. First, it is vanishing as the Newton constant tends to zero, secondly,
it approaches zero as the vacuum expectation value of the scalar field in the YMH theory, or the
pion decay constant in the Skyrme model, becomes zero.

There are also hairy black holes  supporting the Yang-Mills-Higgs hair of the monopole-antimonopole pairs,
chains, and vortex ring solutions with a small black hole placed at the center
\cite{Kleihaus:2000kv,Ibadov:2005rb}. These static
black holes possess non-trivial non-Abelian magnetic field outside their regular event horizon, furthermore, they
provide a counter-example to the black holes uniqueness theorem \cite{Lee:1991vy,Breitenlohner:1994di}.
Similar to the case of the hairy black holes in the Einstein-Skyrme model,
there are two branches of solutions, which emerge from the two globally regular
solutions and  bifurcate at a maximal value of the horizon size
\cite{Lee:1991vy,Breitenlohner:1994di,Kleihaus:2000kv,Ibadov:2005rb}. In a contrary, the  Einstein-Yang-Mills
hairy black hole solutions exist for arbitrarily large horizon size
\cite{Volkov:1989fi,Volkov:1990sva,Kleihaus:1997ic,Bizon:1990sr,Kuenzle:1990is}.

Following the strategy of \cite{Kleihaus:2000kv,Ibadov:2005rb} we shall, in this paper, study the existence of
a new type of hairy black hole solutions, which
correspond to the static axially symmetric Skyrmion-anti-Skyrmion chains with an event horizon at the center.
For the chains with an odd number of constituents a small black hole is placed into the central Skyrmion,
such configuration can be considered as a deformation of the usual spherically symmetric
black hole with Skyrmion hairs. Similarly, for the chains with even number of components,
a black hole is immersed in-between the central Skyrmion-anti-Skyrmion pair.

This paper is organised as follows. In Section II we describe the  Einstein-Skyrme model and construction of the
axially-symmetric solutions representing chains of interpolating Skyrmion-anti-Skyrmion.
We restrict our consideration to the configuration with constituents of
topological degree one and two and consider the configurations with number of components
$k\le 3$. We found that the solutions possess a branch structure for their global quantities both in
terms of the gravitational coupling and in event horizon radius.
Numerical results are presented in Section III,
while the conclusions and remarks are formulated in the last Section.

\section{ The model and field equations}

We consider the Einstein-Skyrme theory in asymptotically flat 3+1 dimensional
space. The action of the model reads
\begin{eqnarray}
S=\int d^4x \sqrt{-g} \left[
\frac{R}{16 \pi G} +L_{Sk} \right],
\label{act}
\end{eqnarray}
where $R$ is the curvature scalar, $g$ denotes the metric determinant and $G$ represents Newton's constant.
The matter part of the action $L_{Sk}$ chosen as the Skyrme model
\be
L_{Sk}= \frac{F_\pi^2}{16}\; g^{\mu\nu} \Tr \left( L_\mu L_\nu \right)
+ \frac{1}{32 e^2}\; g^{\mu\nu} g^{\rho\sigma}
\left[L_\mu,L_\rho\right]\left[L_\nu,L_\sigma\right]
+ \frac{\mu_\pi^2 F_\pi^2}{8} \Tr (U-1)
\ee
with a potential term with a mass parameter $\mu_\pi^2$. Here
$F_\pi$ and $e$ are positive coupling constants and
\be
\label{cur}
L_\mu=U^\dagger \partial_\mu U
\ee
is the $\mathfrak{su}(2)$-valued left-invariant current, associated with
the $\mbox{SU}(2)$-valued scalar field $U=\sigma \cdot {\mathbb{I}} + i \pmb{\pi} \cdot \pmb{\tau}$.
It can be represented in terms of the quartet of
scalar fields $\phi^a=\left(\sigma,\pmb{\pi}\right)$ restricted to the surface of the unit sphere $S^3$,
$(\phi^a)^2 = \sigma^2+\pmb{\pi}\cdot\pmb{\pi}=1$.

The field of the model is required to satisfy the boundary condition $U({\bf x}) \to \mathbb{I}$
as ${\bf x} \to \infty$,  the field
is a map $U: S^3 \mapsto S^3$ labeled by the topological invariant $B=\pi_3(S^3)$. Explicitly,
\be
\label{topcharge}
B=\frac{1}{24\pi^2}\int d^3x \varepsilon^{ijk}~\tr \left[(U^\dagger \partial_i U)
(U^\dagger \partial_j U)
(U^\dagger \partial_k U)\right] = \int d^3x |g|^{1/2} B_0
\ee
where $B_0$ is the temporal component of the topological current
\be
\label{topcur}
B^\mu=\frac{1}{24\pi^2|g|^{1/2}}\varepsilon^{\mu\nu\rho\sigma}\Tr\left(L_\nu L_\rho L_\sigma \right).
\ee

We note that a rescaling of the radial coordinate $r\to erF_\pi/2$ transforms the action of the
Einstein-Skyrme model \re{act} to the form
\be
S=\int d^4x \sqrt{-g} \left\{
\frac{R}{\alpha^2} +\frac12
g^{\mu\nu} \Tr \left( L_\mu L_\nu \right)
+ \frac{1}{16}\; g^{\mu\nu} g^{\rho\sigma}
\Tr \left(\left[L_\mu,L_\rho\right]\left[L_\nu,L_\sigma\right]\right)
+ \mu^2 \Tr (U-1)
 \right\} \, ,
\label{act-scale}
\ee
where $\mu=2\mu_\pi/(eF_\pi)$ is the rescaled mass parameter and $\alpha^2 = 4\pi GF_\pi$
is the effective gravitational coupling. In our numerical simulations we set  $\mu=1$.

Variation of the rescaled action \re{act-scale} with respect
to the metric leads to the Einstein
equations
\be
R_{\mu\nu}-\frac12 R g_{\mu\nu} = \alpha^2 T_{\mu\nu}
\label{Eeq}
\ee
where the Skyrme stress-energy tensor is
\be
\begin{split}
T_{\mu\nu}&= \Tr\left( \frac12 g_{\mu\nu} L^\alpha L_\alpha - L_\mu L_\nu\right)
+\Tr\left( g_{\mu\nu} [L_\alpha, L_\beta][L^\alpha,L^\beta]
- 4 g_{\alpha\beta}[L_\mu, L^\alpha][L_\nu,L^\beta]\right)\\
&+\mu^2 g_{\mu\nu}\Tr (U-1)\, .
\end{split}
\ee

Both the regular self-gravitating Skyrmions and the static axially symmetric black hole solutions
can be constructed in isotropic coordinates with the Lewis-Papapetrou metric
\be
ds^2=-fdt^2+\frac{m}{f}\left(dr^2+r^2 d\theta^2\right)+\frac{l}{f}r^2\sin^2\theta d\varphi^2,
\label{metric}
\ee
where the metric functions
$f$, $m$ and $l$, are functions of
the radial variable $r$ and polar angle $\theta$, only.

Note that,making use of the axially symmetric  parametrization of the Skyrme fields
\be
\label{field}
\pi_1= \phi_1 \cos(n \varphi);\quad \pi_2=\phi_1 \sin(n \varphi);
\quad \pi_3=\phi_2;\quad \sigma = \phi_3
\ee
where $n\in \mathbb{Z}$ is the azimuthal winding number and
$\phi_a$ is a triplet of field variables on the unit sphere $S^2$, we can implement
so-called "trigonometric" parametrization, see e.q. \cite{Sawado:2004yq}
\be
\pi_1 = \sin P \sin Q\, ;\quad
\pi_2 = \sin P \cos Q\, ;\quad
\pi_3 = \cos P\, ,
\label{trig}
\ee
where the Skyrmion's profile functions $P(r,\theta)$ and $Q(r,\theta)$ depend on the radial
coordinate $r$ and polar angle $\theta$. The value of the topological charge \re{topcharge} depends on the
boundary conditions on these two functions \cite{Krusch:2004uf,Shnir:2009ct,Shnir:2015aba}. Imposing
\be
\lim_{r\to \infty} Q(r,\theta)= k\theta\, ,
\ee
where the second integer $k$ specifies the asymptotic value of the field $Q(r,\theta)$,
we obtain $B=\frac{n}{2}\left(1-(-1)^k\right)$. Thus, the case $k=1$ corresponds to the multi-Skyrmions of
topological degree $B=n$, while $k=2$ yields the Skyrmion-anti-Skyrmion (S-A)
static sphaleron solution of the Einstein-Skyrme model, consisting of a charge $n$
Skyrmion and a charge $-n$ anti-Skyrmion \cite{Krusch:2004uf,Shnir:2009ct,Shnir:2015aba}.
Configurations with $k\ge 3$ correspond to the Skyrmion-anti-Skyrmion chains with $k$ constituents
placed along the axis of symmetry in alternating order.

\section{Results}
\subsection{Boundary conditions}

To obtain asymptotically flat solutions of the Einstein-Skyrme equations,
which are either globally regular or possess a
regular event horizon, we must impose appropriate boundary
conditions \cite{Luckock:1986tr,Droz:1991cx,Bizon:1992gb,Ioannidou:2006nn,Shnir:2015aba,Sawado:2004yq}.

Here we consider static axially symmetric black hole solutions with "Skyrmion-anti-Skyrmion hair", which are
asymptotically flat, and possess a finite mass. Then the corresponding
boundary conditions at spatial infinity and along the symmetry axis are the same as those of the regular
self-gravitating Skyrmions \cite{Bizon:1992gb,Shnir:2015aba,Sawado:2004yq}. In particular,
as $r\to \infty$, the asymptotic value of the Skyrme
field is restricted to the vacuum and the metric functions must approach unity.
Explicitly, we impose
\be
\label{BCinf}
\begin{split}
\phi_1\bigl.\bigr|_{r\rightarrow\infty}&\rightarrow 0,\quad \phi_2\bigl.\bigr|_{r\rightarrow\infty}\rightarrow 0,\quad
\phi_3\bigl.\bigr|_{r\rightarrow\infty}\rightarrow 1,\quad \\
f\bigl.\bigr|_{r\rightarrow\infty}&\rightarrow 1,\quad l\bigl.\bigr|_{r\rightarrow\infty}\rightarrow 1,\quad
m\bigl.\bigr|_{r\rightarrow\infty}\rightarrow 1 \, .
\end{split}
\ee
The condition of regularity of the functions on the symmetry axis yields
\be
\label{BCz}
\begin{split}
\phi_1\bigl.\bigr|_{\theta=0}&=0,\quad \partial_\theta \phi_2\bigl.\bigr|_{\theta=0}=0,
\quad \partial_\theta \phi_3\bigl.\bigr|_{\theta=0}=0,\quad \\
\partial_{\theta} f\bigl.\bigr|_{\theta=0}&=0,\quad \partial_{\theta} l\bigl.\bigr|_{\theta=0}=0,\quad
\partial_{\theta} m\bigl.\bigr|_{\theta=0}=0\, .
\end{split}
\ee
We further impose, that the two metric functions $m(r,\theta), l(r,\theta)$ on the symmetry
axis satisfy
$$
m(r,\theta=0,\pi)=l(r,\theta=0,\pi)
$$
This condition secures the absence of a conical singularity, it
requires that the deficit angle should vanish.

Note that the non-linear system of the Einstein-Skyrme equations includes three equations on the matter
fields $\phi_a$. One can try to reduce the number of equations, considering the trigonometric
parametrization of the triplet $\phi_a$ given by ansatz \re{trig}
\cite{Sawado:2004yq}. However, in such a case, there are obstacles related with
regularity conditions we have to impose on the angular function $Q(r,\theta)$ for $k\ge 2$
\cite{Shnir:2009ct}, thus we just make use of the parametrization \re{trig} to generate an appropriate input
configuration with $k$ components.

To obtain globally regular solutions, we must impose appropriate boundary conditions at the origin.
For odd values of the integer $k$ the boundary conditions are identical to those for the
case of the spherically symmetric fundamental gravitating Skyrmion
\be
\label{BCor-odd}
\begin{split}
\phi_1\bigl.\bigr|_{r\rightarrow 0}&\rightarrow 0,\quad
\phi_2\bigl.\bigr|_{r\rightarrow 0}\rightarrow 0,\quad
\phi_3\bigl.\bigr|_{r\rightarrow 0}\rightarrow -1,\quad \\
\partial_r f\bigl.\bigr|_{r\rightarrow 0}&\rightarrow 0,\quad
\partial_r l\bigl.\bigr|_{r\rightarrow 0}\rightarrow ,\quad
\partial_r m\bigl.\bigr|_{r\rightarrow 0}\rightarrow 0 \, .
\end{split}
\ee
while for the S-A pair and other Skyrmion-anti-Skyrmion chains with even number of components,
the boundary conditions on the matter field are different
\cite{Krusch:2004uf,Shnir:2009ct,Shnir:2015aba}:
\be
\label{BCor-even}
\phi_1\bigl.\bigr|_{r\rightarrow 0}\rightarrow 0,\quad
\partial_r \phi_2\bigl.\bigr|_{r\rightarrow 0}\rightarrow 0,\quad
\partial_r \phi_3\bigl.\bigr|_{r\rightarrow 0}\rightarrow 0
\ee

The event horizon of the static black hole resides at a surface of constant radial coordinate, $r = r_h$.
The boundary conditions are obtained from the
asymptotic expansion of the corresponding
field equations near the horizon.
Regularity at $r=r_h$ requires that
\be
\label{BC-rh}
\begin{split}
f\bigl.\bigr|_{r\rightarrow r_h}&\rightarrow 0,\quad
m \bigl.\bigr|_{r\rightarrow r_h}\rightarrow 0,\quad
l\bigl.\bigr|_{r\rightarrow r_h}\rightarrow 0\quad \\
\partial_r\phi_1\bigl.\bigr|_{r\rightarrow r_h}&\rightarrow 0,\quad
\partial_r \phi_2\bigl.\bigr|_{r\rightarrow r_h}\rightarrow 0,\quad
\partial_r \phi_3\bigl.\bigr|_{r\rightarrow r_h}\rightarrow 0 \, .
\end{split}
\ee
In our numerical calculations we make use of the parametrization of the metric functions
\be
\label{metric-repar}
f(r,\theta)= f_2(r,\theta) \frac{\left(1-\frac{r_h}{r}\right)^2}{\left(1+\frac{r_h}{r}\right)^2}, \quad
l(r,\theta)= l_2(r, \theta)\left(1-\frac{r_h}{r}\right)^2,\quad
m(r,\theta)= m_2(r,\theta)\left(1-\frac{r_h}{r}\right)^2
\ee
and impose the following boundary conditions at $r=r_h$
$$
\partial_r f_2 \bigl.\bigr|_{r\rightarrow r_h}\rightarrow 0,\quad
\partial_r m_2 \bigl.\bigr|_{r\rightarrow r_h}\rightarrow 0,\quad
\partial_r l_2 \bigl.\bigr|_{r\rightarrow r_h}\rightarrow 0
$$

We have solved the boundary value problem
for the coupled system of six nonlinear partial differential equations
with boundary conditions \re{BCinf}-\re{BC-rh} using
a six-order finite difference scheme.

Within our formulation, the numerical problem possesses
five input parameters: $r_h$, $\alpha$, $n$, $k$  and the mass parameter $\mu$.
The emerging overall system becomes rather complicated
and we did not attempt to explore in a systematic way the entire
parameter space of all solutions. In particular, we restricted our analysis to the configurations with
winding $n=1,2$
and consider only chains with two and three components. We also did not study the dependency of the solutions
on the value of the mass parameter setting $\mu=1$.

To facilitate  the  calculations  in  the  near
horizon  area,  we  have  introduced  the new compact
radial coordinate $x=\frac{r-r_h}{r+c}$, which  maps the semi-infinite region $r\in [r_h,\infty)$
onto the unit interval $x\in [0,1]$. Here $c$ is an arbitrary constant
used to adjust the contraction of the grid.
The system of equations is discretized on a grid
with typical number of points $89\times 69$.
The underlying linear system is solved with the packages FIDISOL/CADSOL \cite{schoen}. The typical
errors are of order of $10^{-4}$.

\subsection{Quantities of interest and horizon properties}
Asymptotic expansions of the metric functions at the
horizon and at spatial infinity yield important physical
properties of the BHs. The total ADM mass of the configuration
can be read of from the asymptotic subleading behavior of the metric functions as $r\to \infty$:
\be
\label{massfinal}
M=\frac{1}{2G}\lim\limits_{r\rightarrow\infty}r^2 \partial_r f \, .
\ee
It is convenient to introduce the rescaled  coordinate $\hat x = x/\alpha$, horizon radius
$\hat r_h =r_h/\alpha$ and rescaled mass $M=\alpha M $.

The  physically  interesting  horizon  properties  include  the
surface gravity
$$
\kappa^2 = -\frac14 g^{00}g^{ij}(\partial_i g_{00})(\partial_j g_{00}) \, .
$$
Taking into account the parametrization \re{metric-repar} and the expansion of the metric
functions in the near-horizon region we obtain the dimensionless surface gravity
\be
\hat \kappa = \kappa/\alpha= \frac{f_2(\theta)}{8 r_h \sqrt{m_2(\theta)}}\, .
\ee
The surface gravity, as well as the Kretschmann scalar $K=R^{\mu\nu\rho\sigma}R_{\mu\nu\rho\sigma}$,
is finite at the event horizon.
The Hawking temperature is proportional to the surface gravity, $T=\frac{\kappa}{2\pi}$, see
e.g. \cite{Wald}. Further, taking into account the parametrization of the metric functions
\re{metric-repar}, we can see that the dimesionless event horizon area is defined as
\be
A=32 \pi r_h^2 \int\limits_0^\pi d\theta \sin\theta  \frac{l_2 m_2}{f_2}\, ,
\label{area}
\ee
it is proportional to the entropy $S=A/4$.

Presence of the axially-symmetric field of the gravitationally bound
Skyrmions deforms the event horizon. As usually, small deformation
is revealed, when measuring the ratio of circumferences of the
horizon along the equator
\be
L_e=16 r_h \int\limits_0^{2\pi}d\varphi \sqrt{\frac{l_2}{f_2}}
\label{Le}
\ee
and along the poles
\be
L_p=32 \pi r_h \int\limits_0^{\pi}d\theta \sqrt{\frac{m_2}{f_2}}
\label{Lp}
\ee

Note that, in the case of the black holes with Skyrme hair,
strictly speaking, the solutions cannot be classified according to the topological mapping \re{topcharge}
between physical and internal spaces. Still, one defines a baryon charge density
performing the integration \re{topcharge} in the exterior region. The event horizon "absorbs" a
part of the baryon charge but the integral  \re{topcharge} never vanishes \cite{Droz:1991cx,Sawado:2004yq}.
As we shall see below, the same holds for the black holes with Skyrmion-anti-Skyrmion hairs.

We construct numerically the black hole solutions for the Skyrmion-anti-Skyrmion pairs
and Skyrmion-anti-Skyrmion-Skyrmion  chains with constituents of degrees one and two.
These asymptotically flat solutions are associated with the regular self-gravitating
solutions discussed in \cite{Shnir:2015aba}. Further, they have many features in common with the
EYMH black holes with magnetic dipole hair \cite{Kleihaus:2000kv} and the static axially
symmetric black hole solutions of the  EYM theory \cite{Kleihaus:2007vf}. In particular,
for a fixed value of the effective gravitational coupling $0\le \alpha \le \alpha_{cr}$,
in both cases
there are two branches of solutions, which are linked to the corresponding flat space sphaleron
configurations and the generalized Bartnik-McKinnon solutions, respectively.
The difference is that, as the gravitational interaction remains relatively weak, the non-Abelian interaction
between the constituents of the EYM is stronger than the dipole-dipole interaction between the Skyrmions
\cite{Krusch:2004uf,Shnir:2009ct}. As a results, the Skyrmion-anti-Skyrmion chains may exist in the
flat space only when each of the constituents carries charge larger than two. As the gravitational attraction
becomes stronger, the S-A chains with constituents of unit charge arise forming the lower branch of solutions,
which may be not linked to the flat space \cite{Shnir:2015aba}.

\begin{figure}[t]
\begin{center}
\includegraphics[width=.26\textwidth, angle=-0]{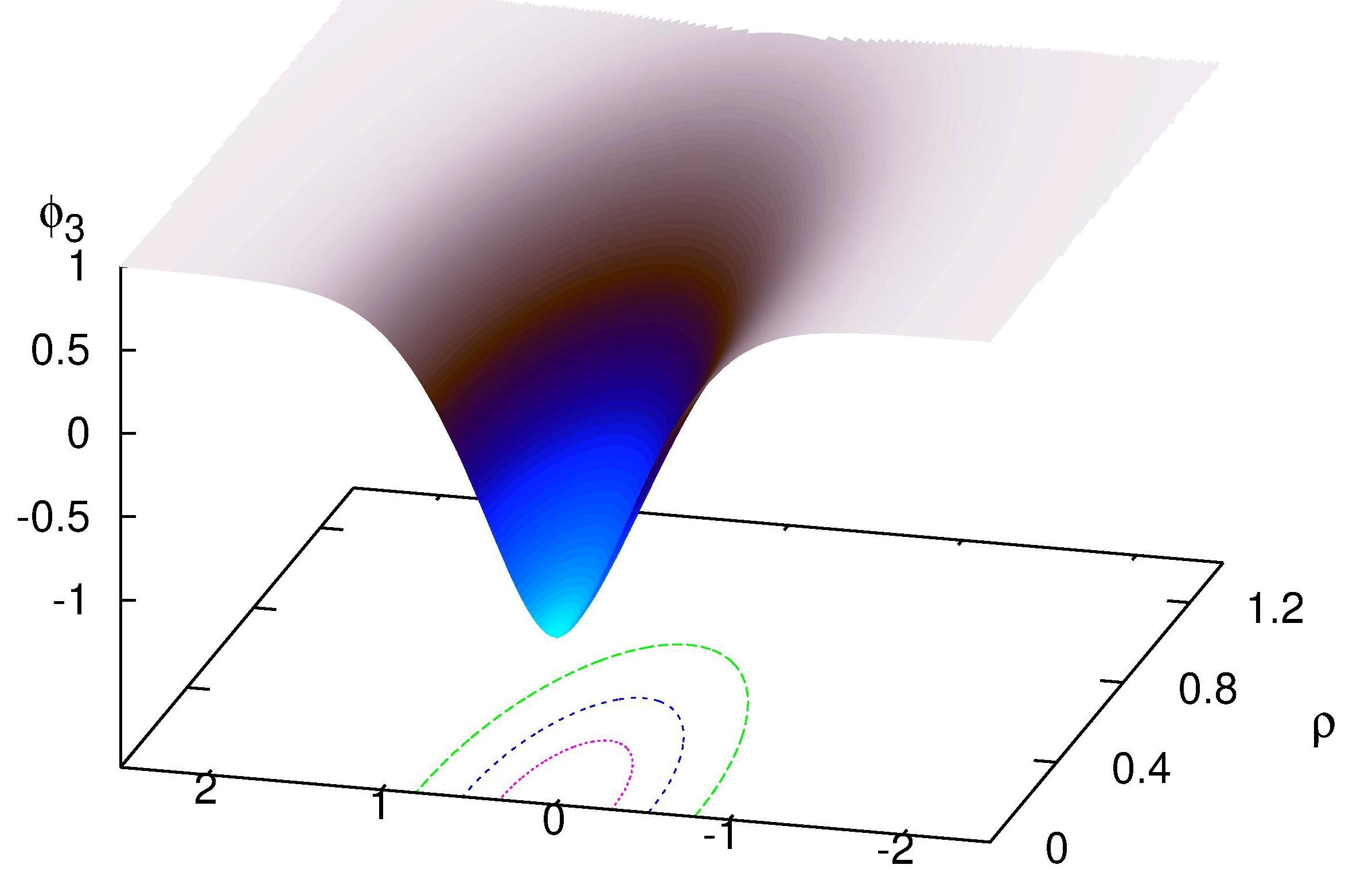}
\includegraphics[width=.26\textwidth,  angle=-0]{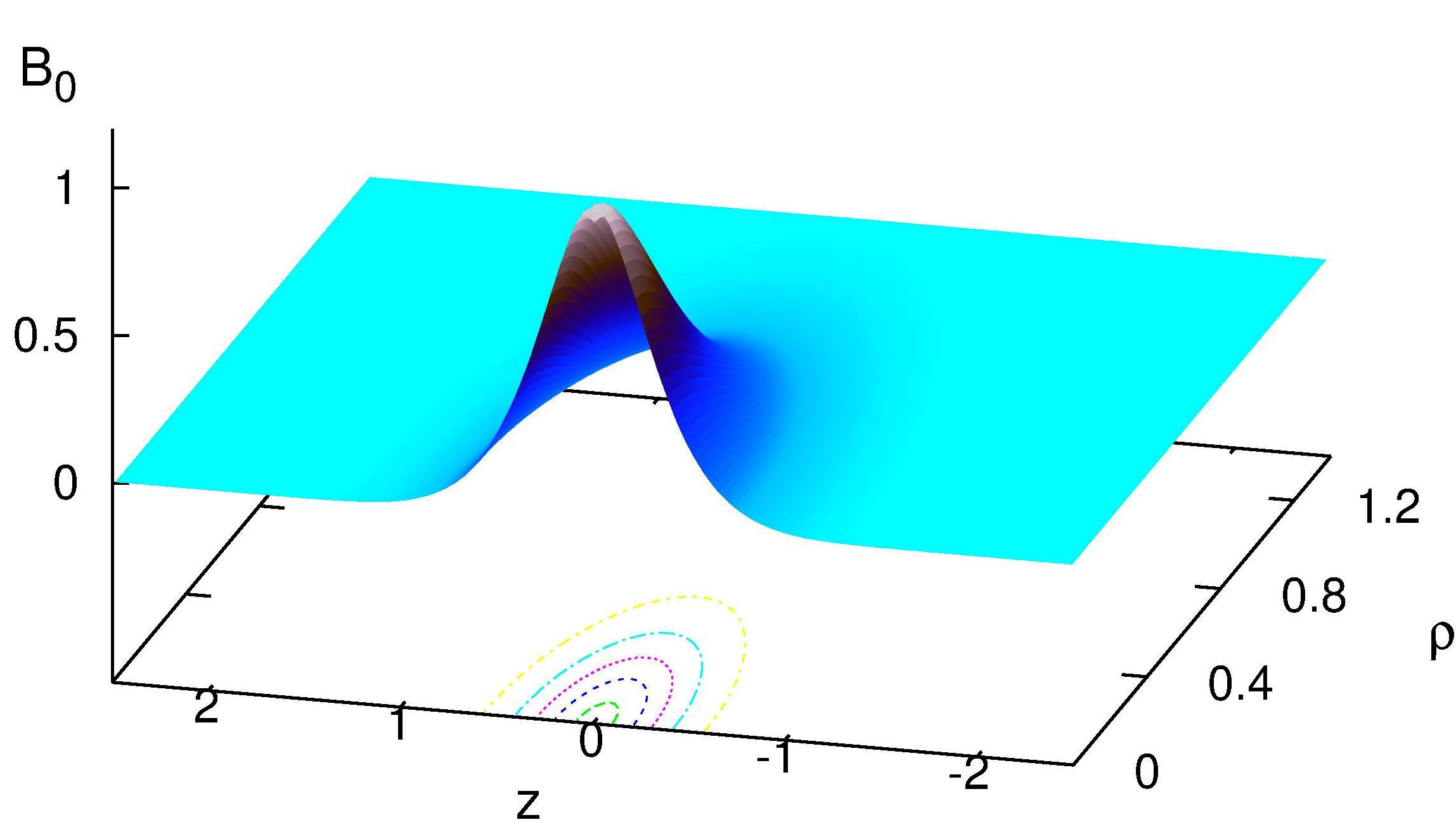}
\includegraphics[width=.26\textwidth,  angle=-0]{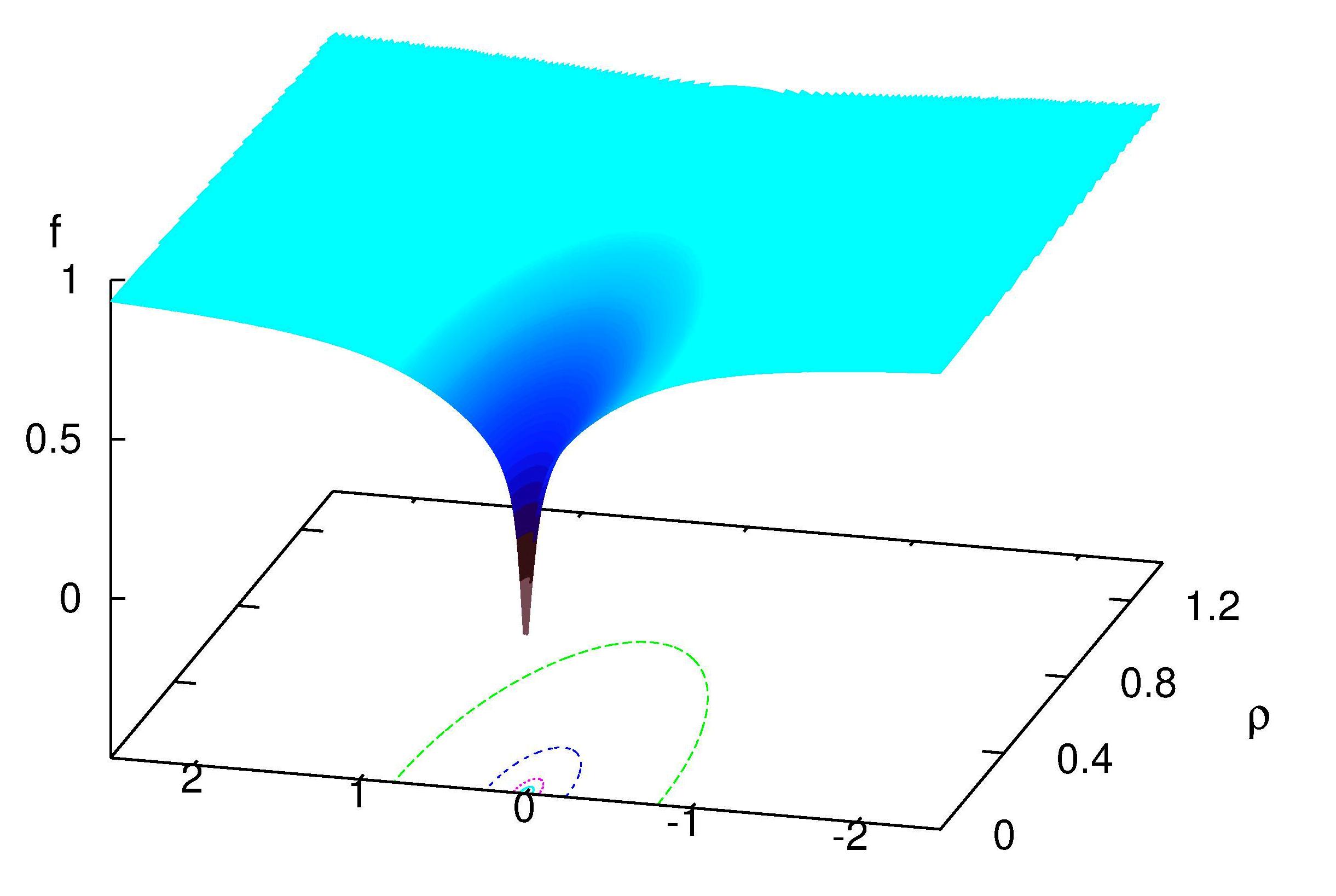}
\includegraphics[width=.26\textwidth, angle=-0]{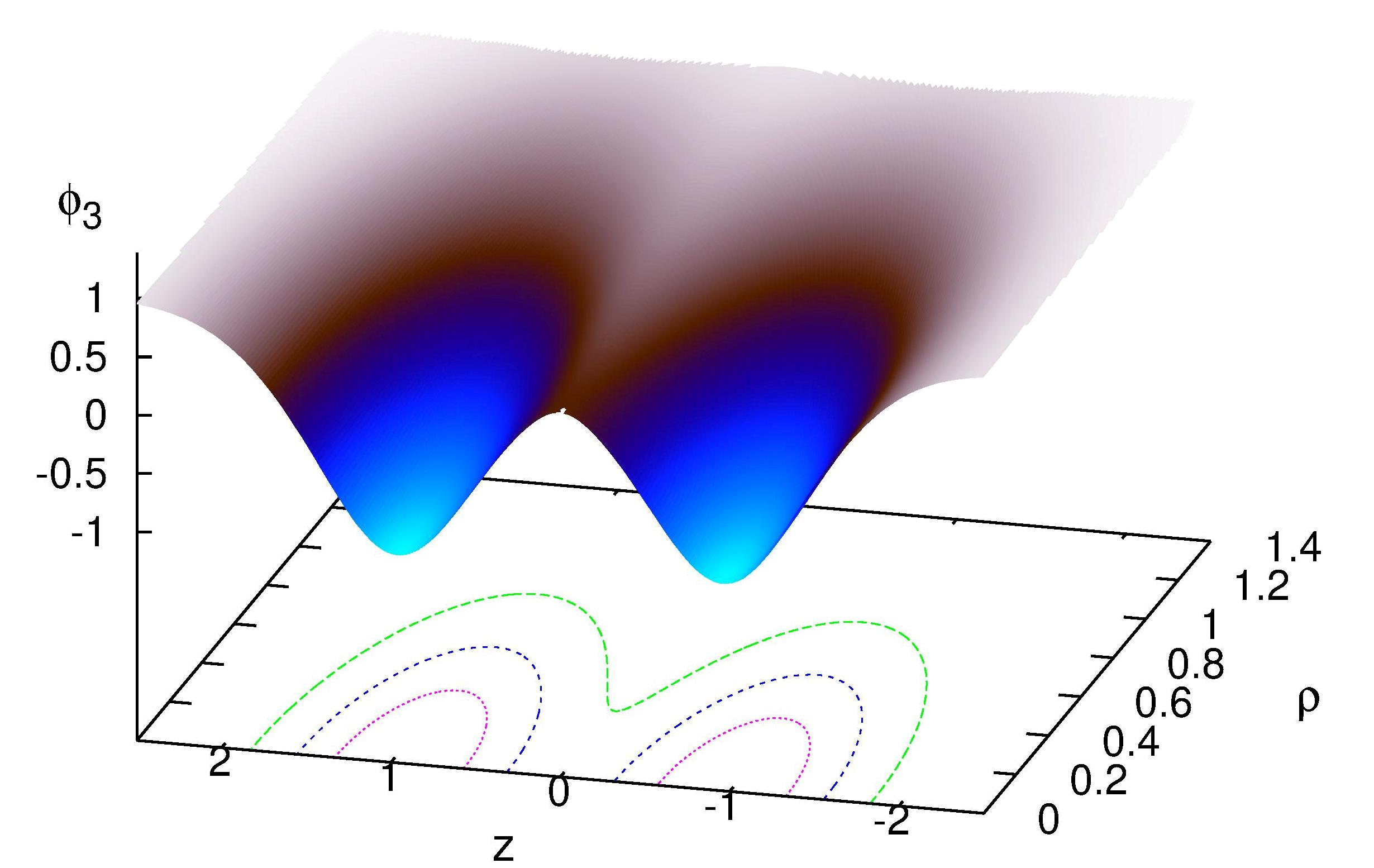}
\includegraphics[width=.26\textwidth, angle=-0]{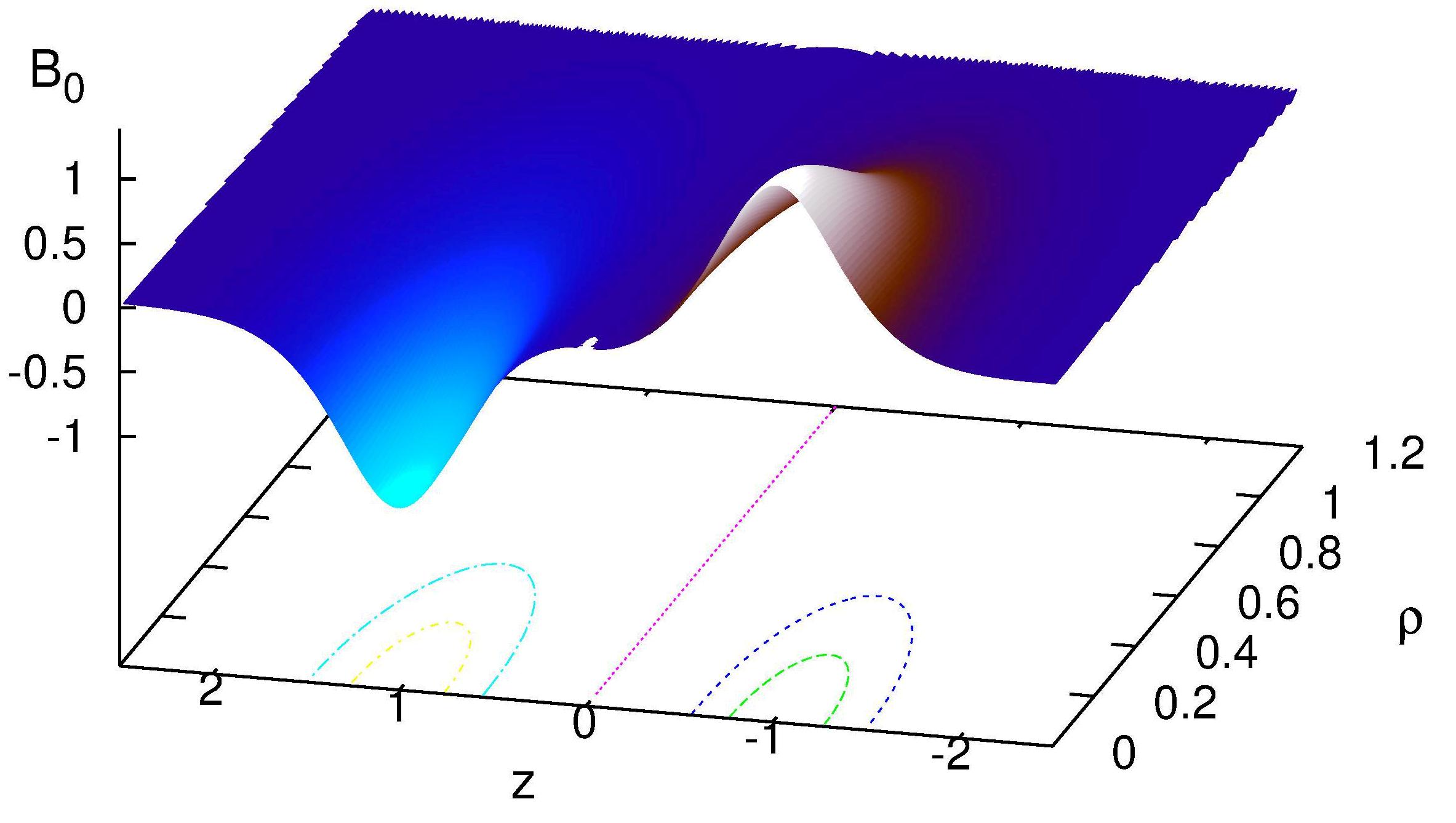}
\includegraphics[width=.26\textwidth,  angle=-0]{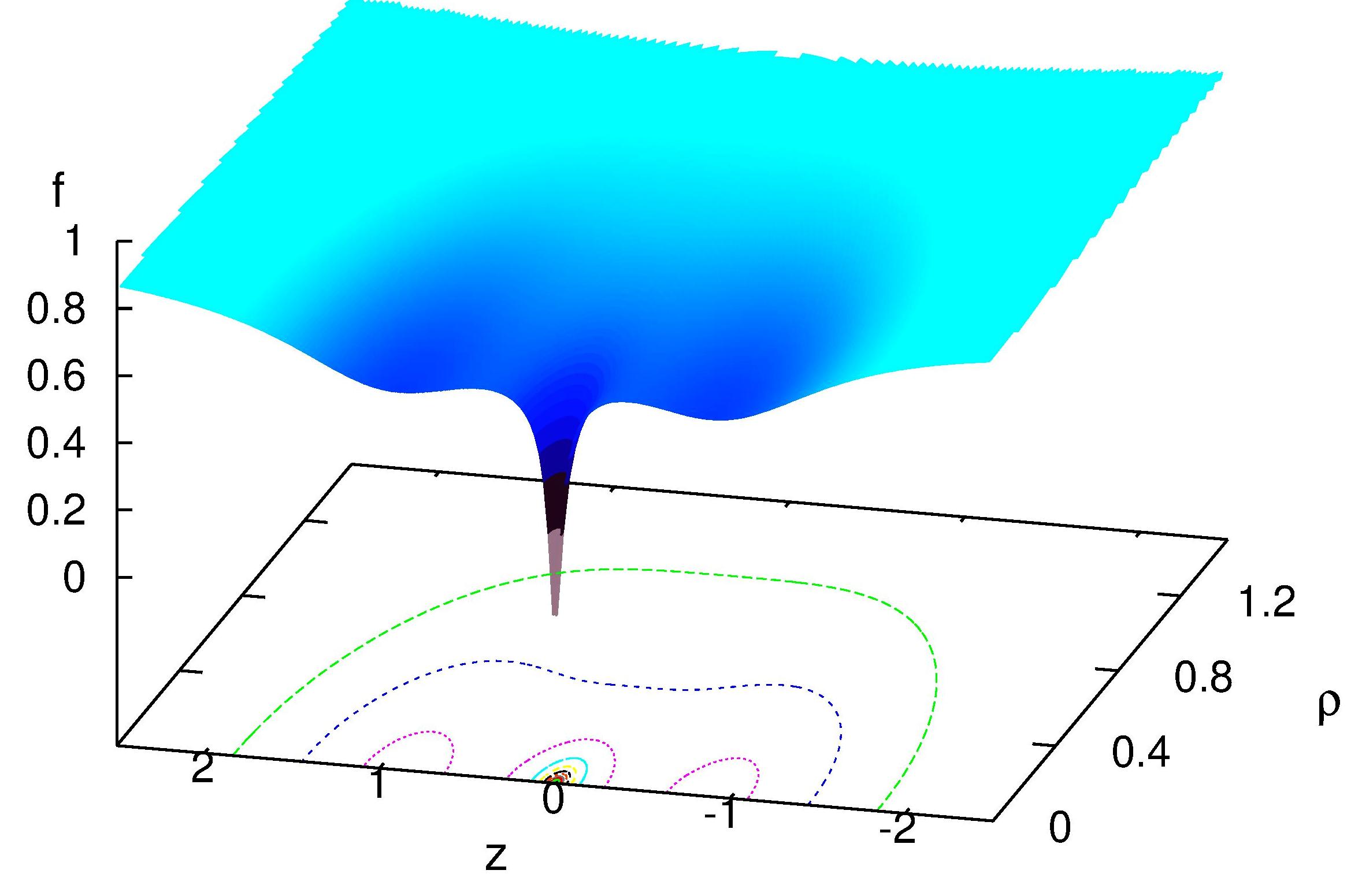}
\includegraphics[width=.26\textwidth,  angle=-0]{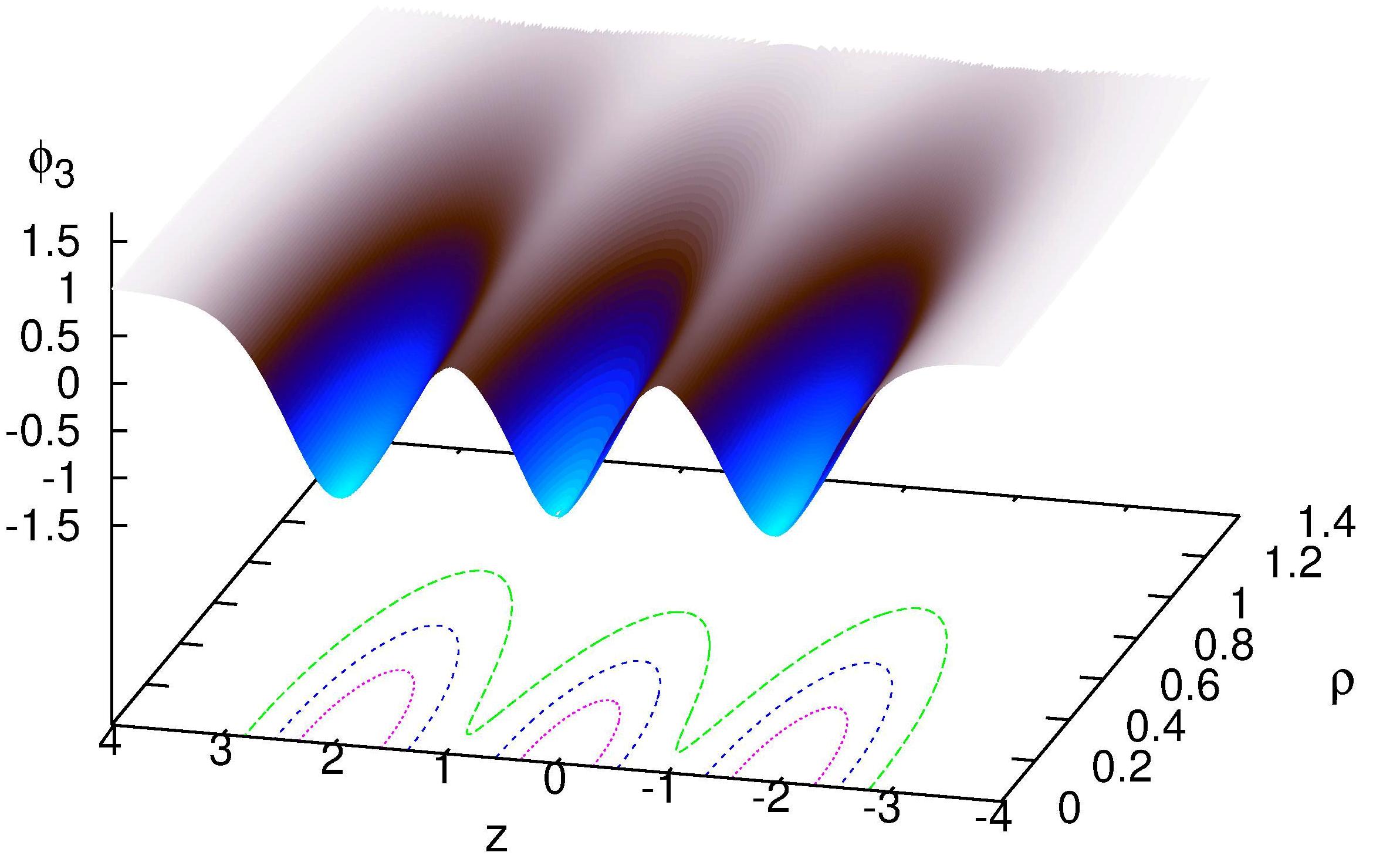}
\includegraphics[width=.26\textwidth,  angle=-0]{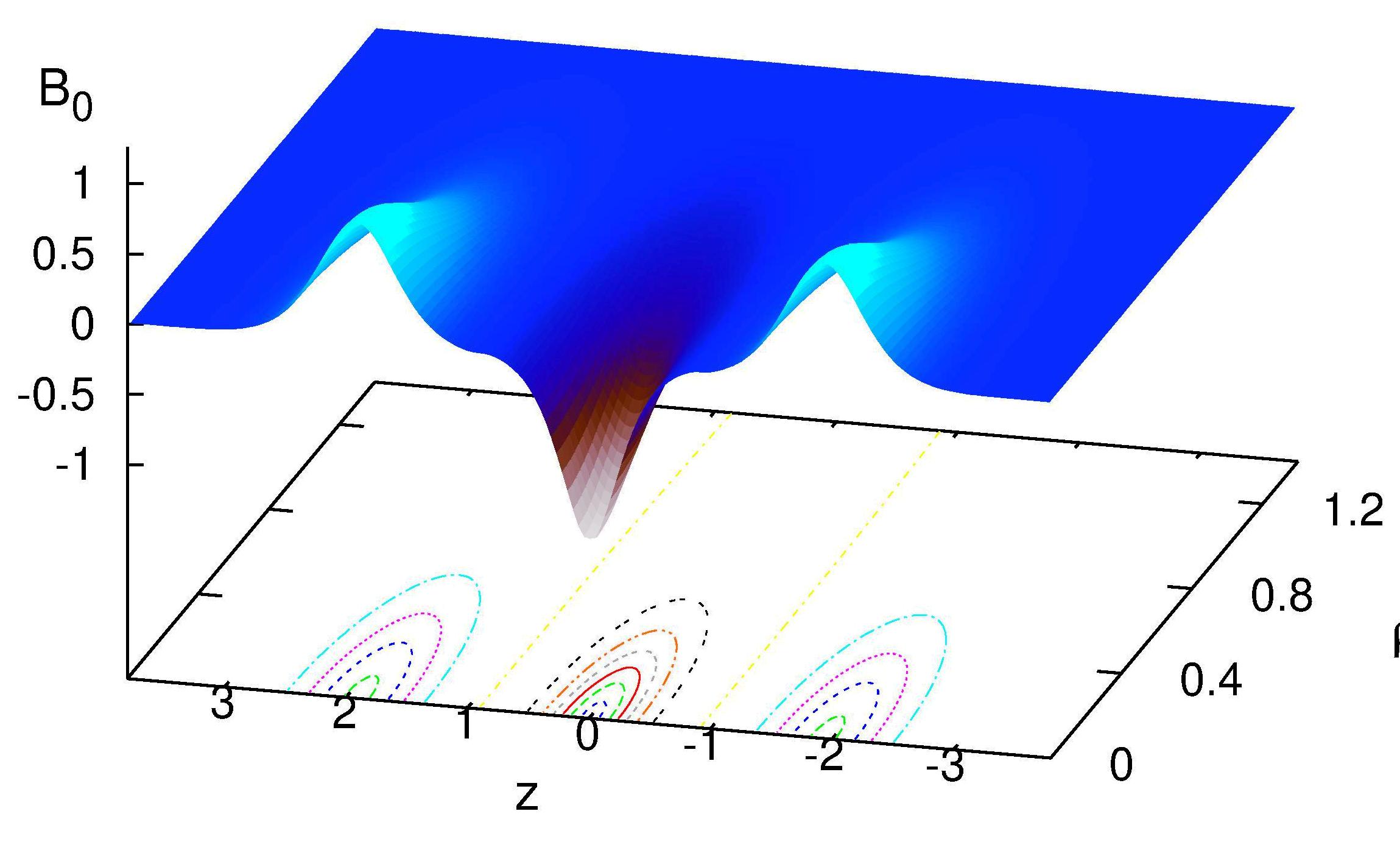}
\includegraphics[width=.26\textwidth,  angle=-0]{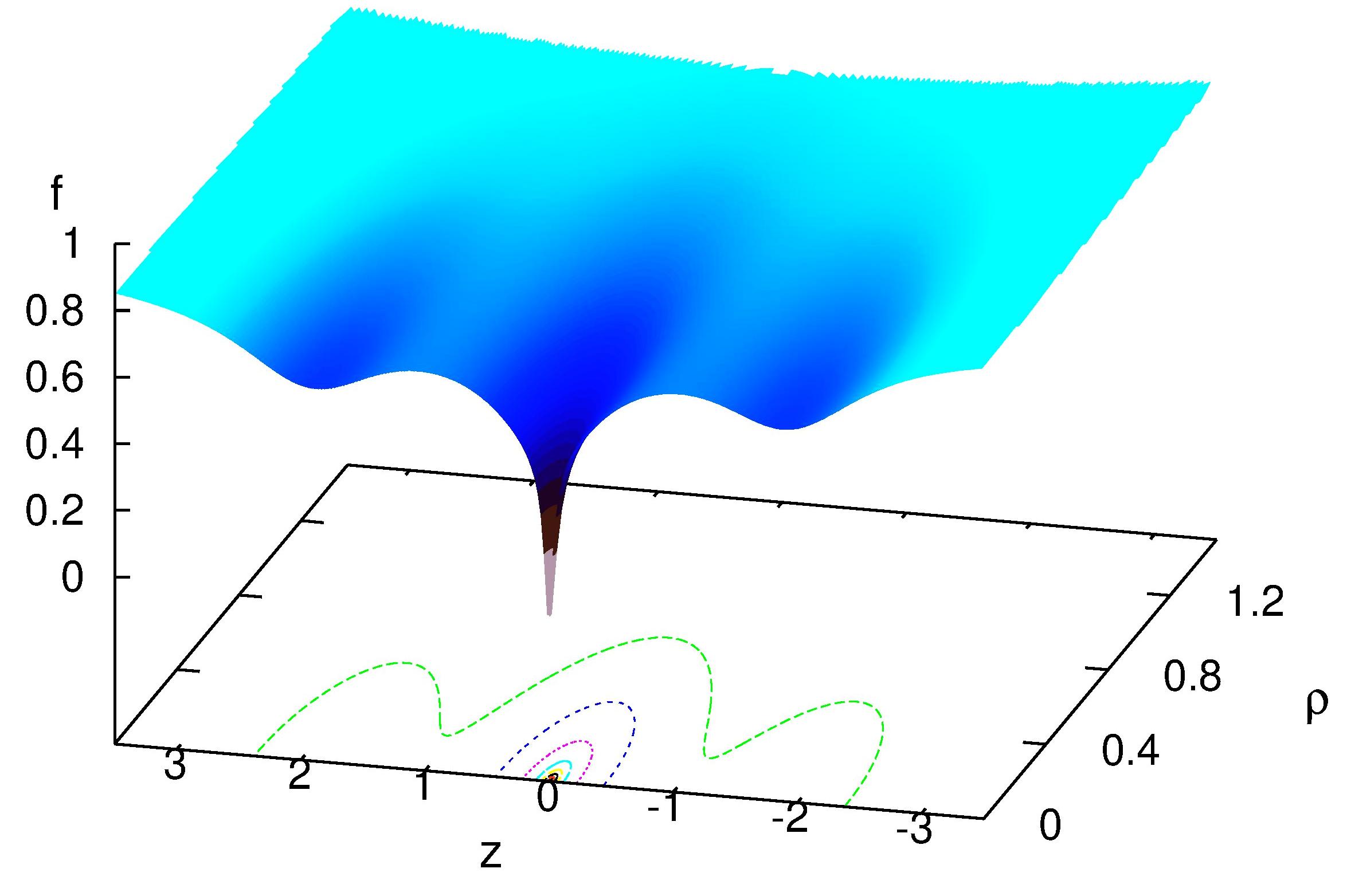}
\end{center}
\caption{\small
Component $\phi_3$ of the Skyrmion field of the hairy BH solutions (left column),
the baryon charge density distribution (middle column) and the
metric function $f$ (right column) of the BHs with Skyrmion-anti-Skyrmion hairs
are plotted for the single black hole with Skyrmion hair (upper row),
the S-A pair (middle row) and for the S-A-S chain (bottom row) on the lower branches of solutions
as functions of the coordinates $z= r \cos \theta$ and
$\rho=r \sin \theta$
at $\alpha=0.15$, $|n|=1$ and $r_h=0.01$.}
\lbfig{fig1}
\end{figure}

\begin{figure}[t]
\begin{center}
\includegraphics[width=.32\textwidth, angle=-90]{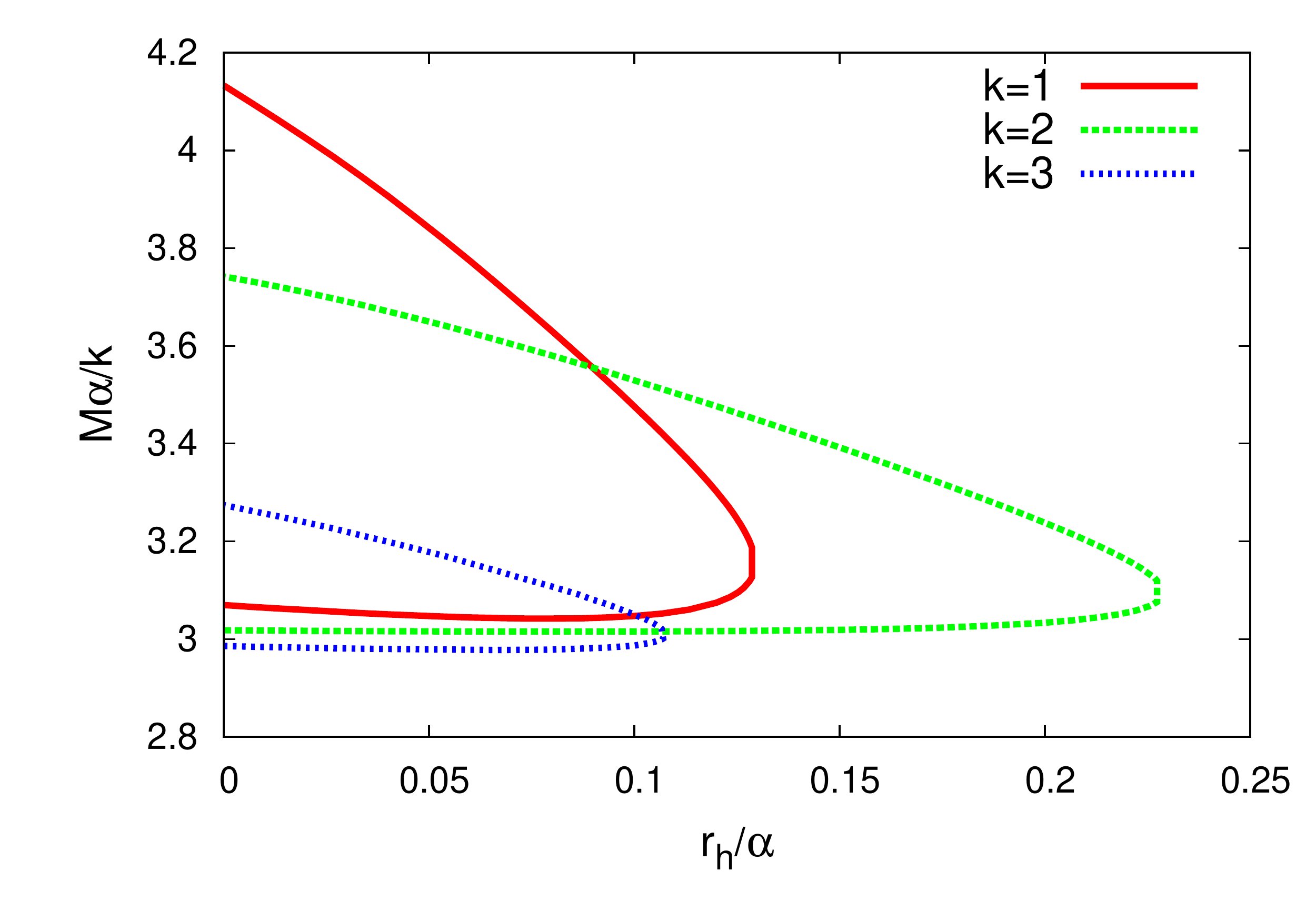}
\includegraphics[width=.32\textwidth, angle=-90]{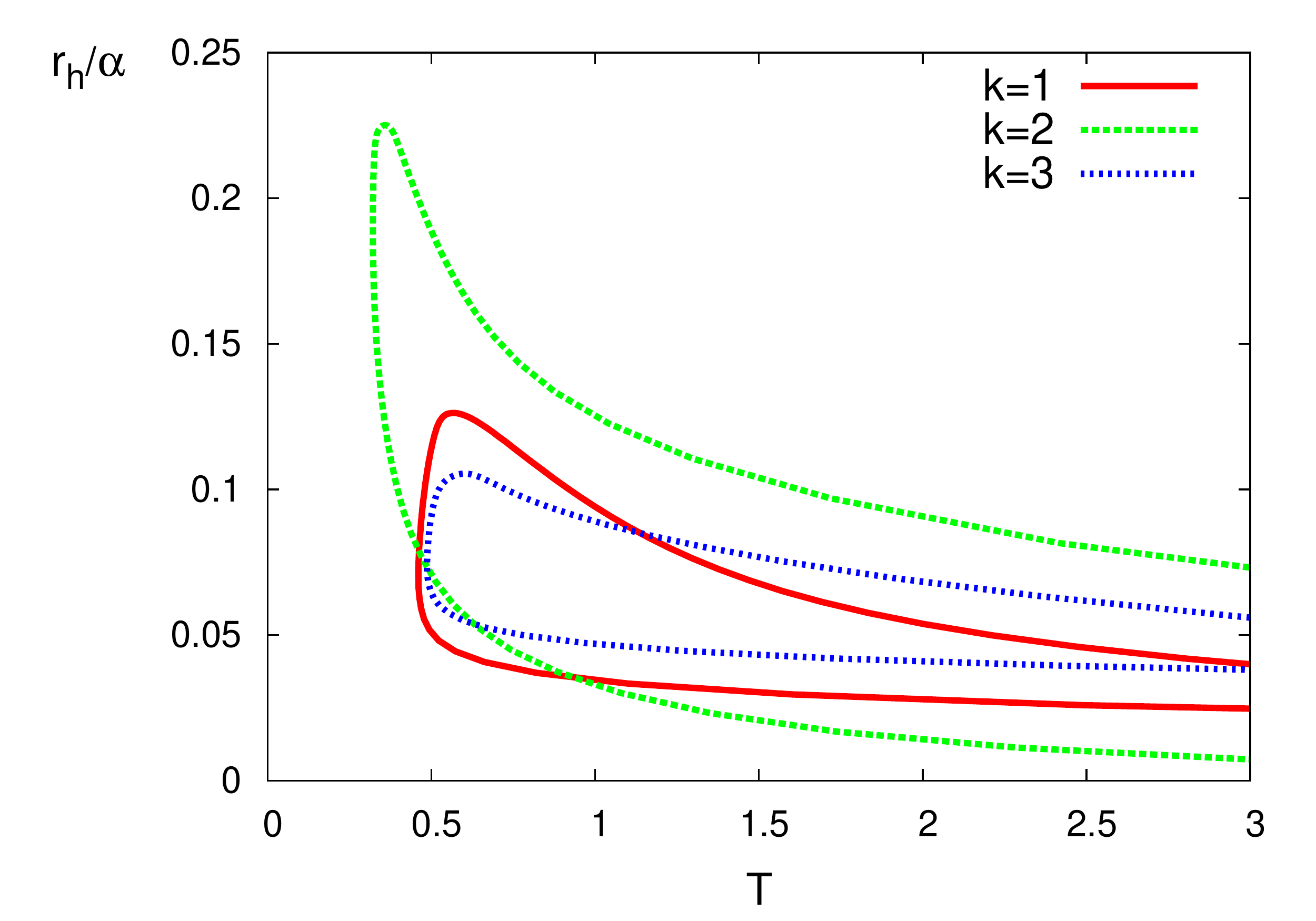}
\end{center}
\caption{\small
The scaled mass $M\alpha$ (divided by the  number of components $k$) of the
static axially symmetric $|n|=1$ Skyrmion bound states (left plot) and the
Hawking temperature (right plot) are shown as functions of the scaled event horizon
radius $r_h/\alpha$ for the single black hole with Skyrmion hair ($k=1$),
the S-A pair ($k=2$) and for the S-A-S chain ($k=3$) at $\alpha=0.15$}.
\lbfig{fig2}
\end{figure}

\begin{figure}[t]
\begin{center}
\includegraphics[width=.32\textwidth, angle=-90]{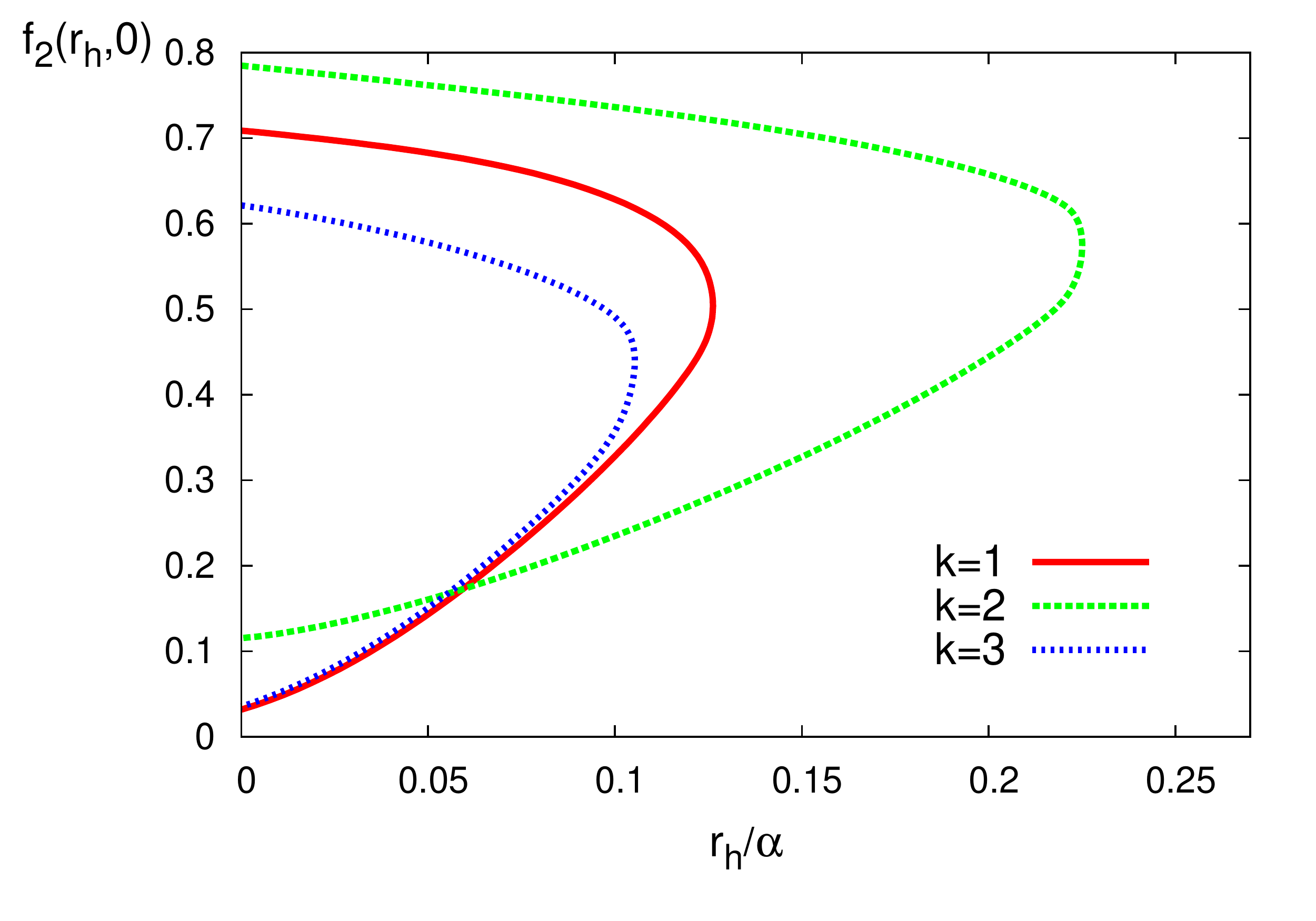}
\includegraphics[width=.32\textwidth, angle=-90]{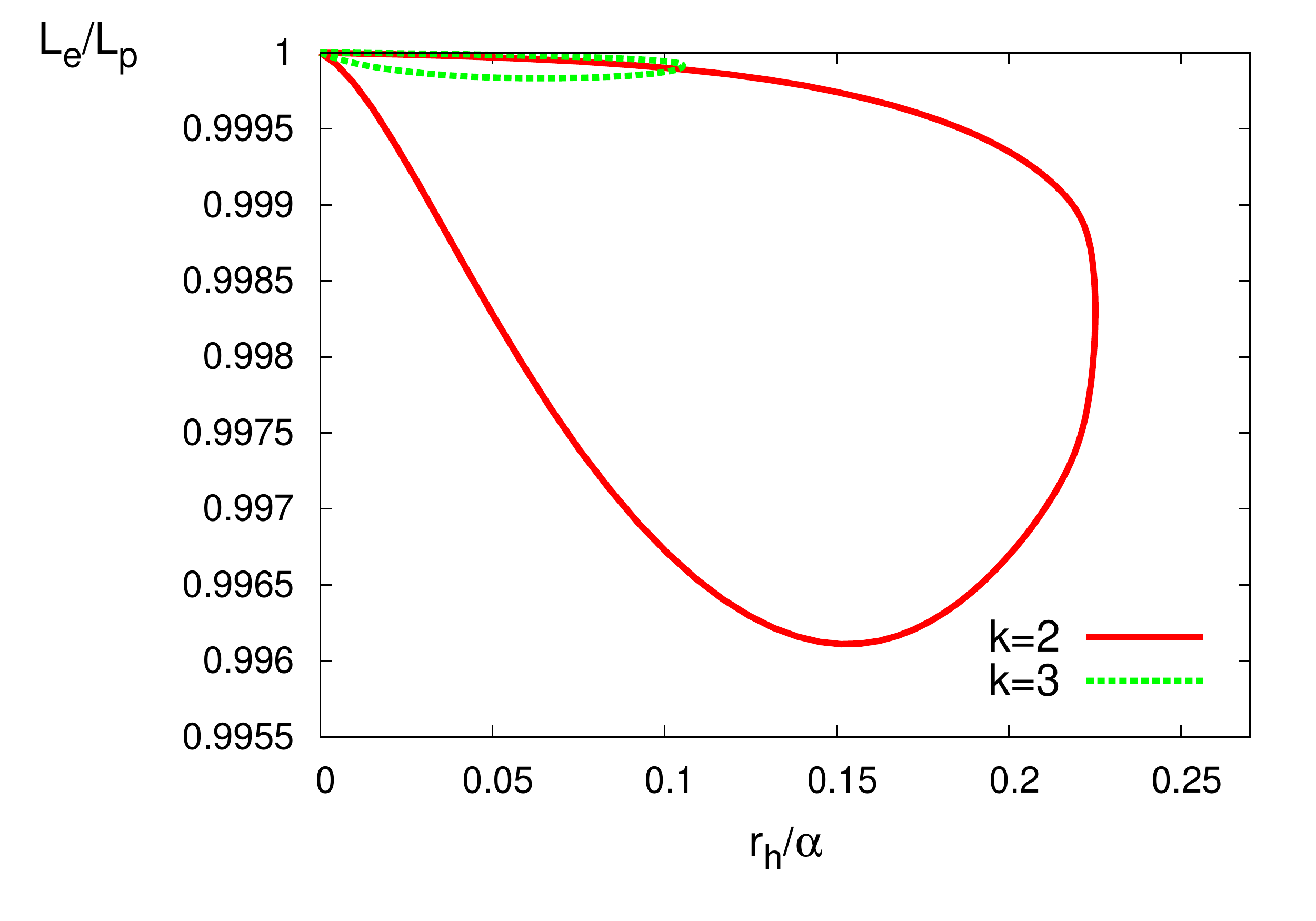}
\end{center}
\caption{\small
The value of the metric function $f_2$ on the event horizon for the
black holes bounded with static $|n|=1$ axially symmetric Skyrmions
(left plot) and the ratio $L_e/L_p$ (which gives a measure of the horizon deformation,
right plot) are shown as functions of the scaled event horizon
radius $r_h/\alpha$ for the single black hole with Skyrmion hair ($k=1$),
the S-A pair ($k=2$) and for the S-A-S chain ($k=3$) at $\alpha=0.15$}.
\lbfig{fig3}
\end{figure}

\begin{figure}[t]
\begin{center}
\includegraphics[width=.32\textwidth, angle=-90]{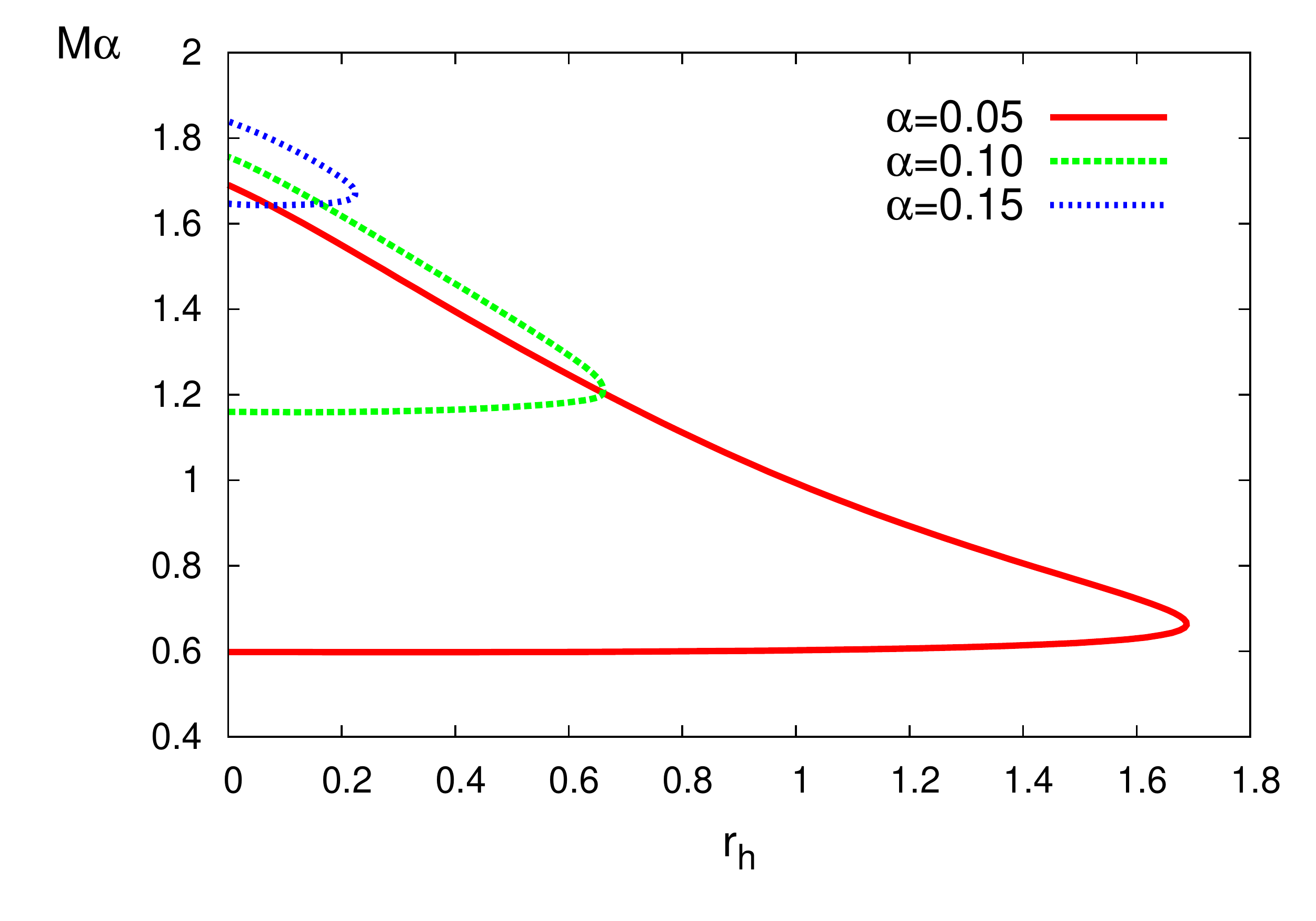}
\includegraphics[width=.32\textwidth, angle=-90]{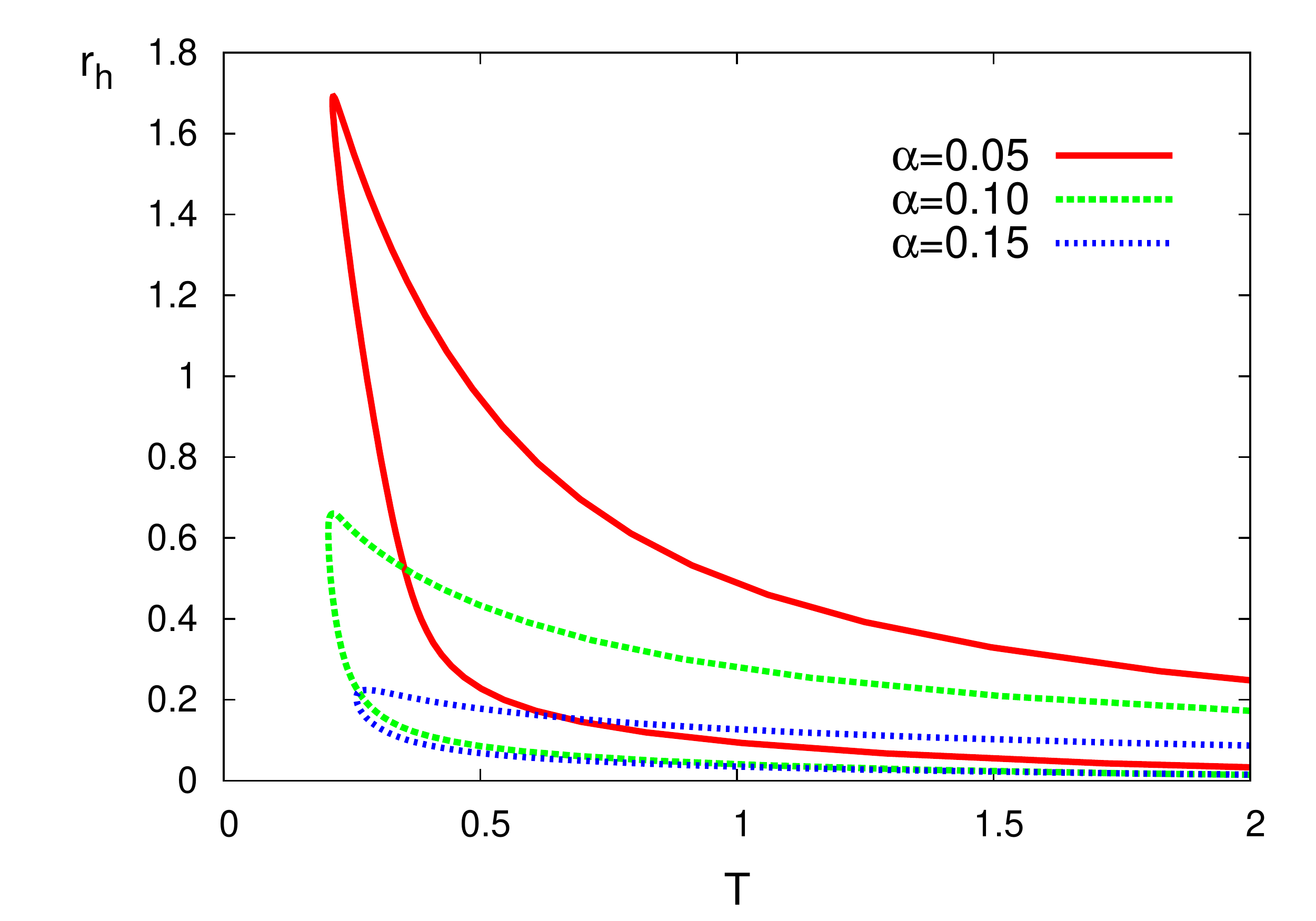}
\end{center}
\caption{\small
The scaled mass $M\alpha$   (left plot) and the
Hawking temperature (right plot) of the $|n|=2$ S-A pair are shown as functions of the scaled event horizon
radius $r_h/\alpha$ for some set of values of  $\alpha$}.
\lbfig{fig4}
\end{figure}

\begin{figure}[t]
\begin{center}
\includegraphics[width=.32\textwidth, angle=-90]{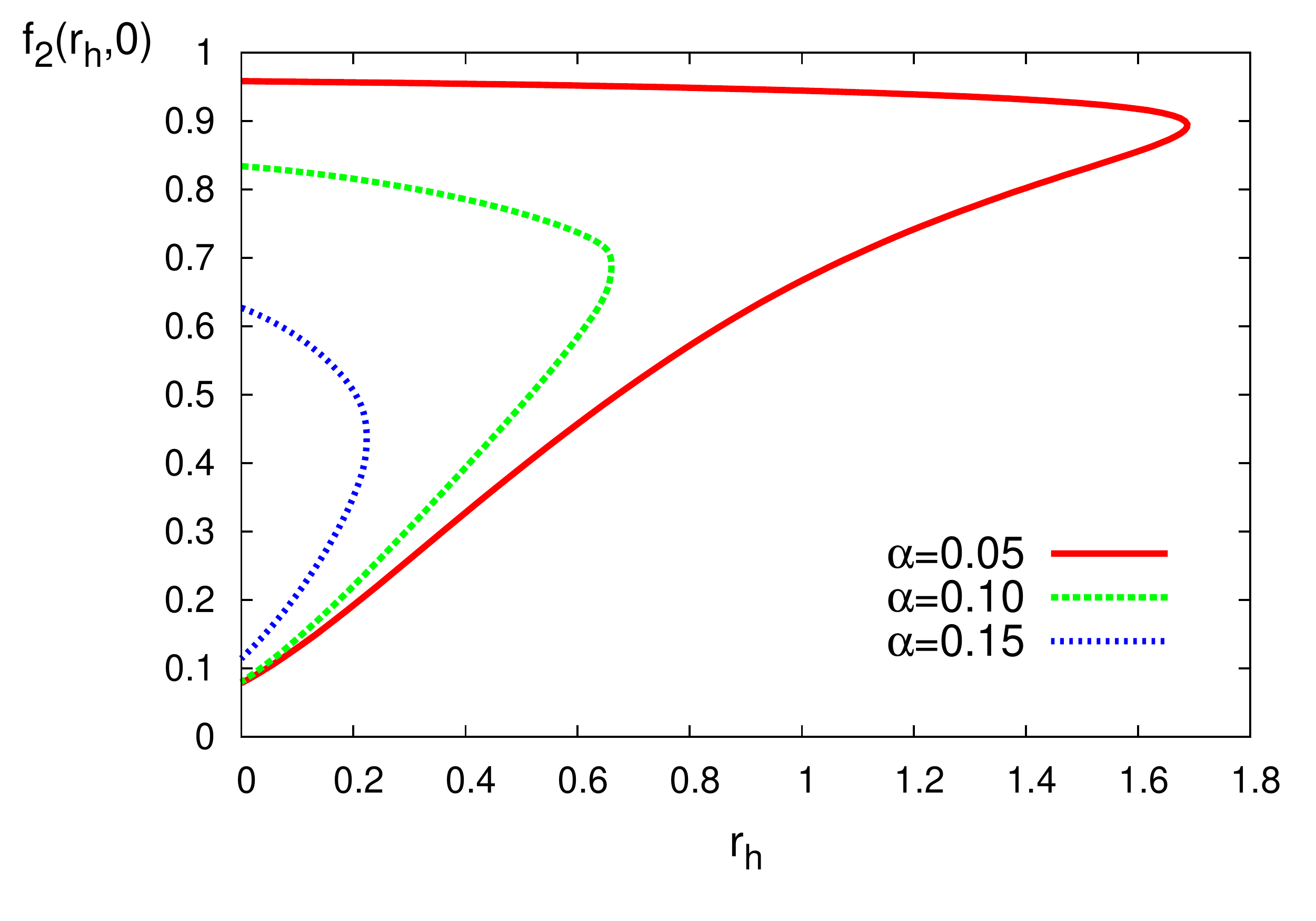}
\includegraphics[width=.32\textwidth, angle=-90]{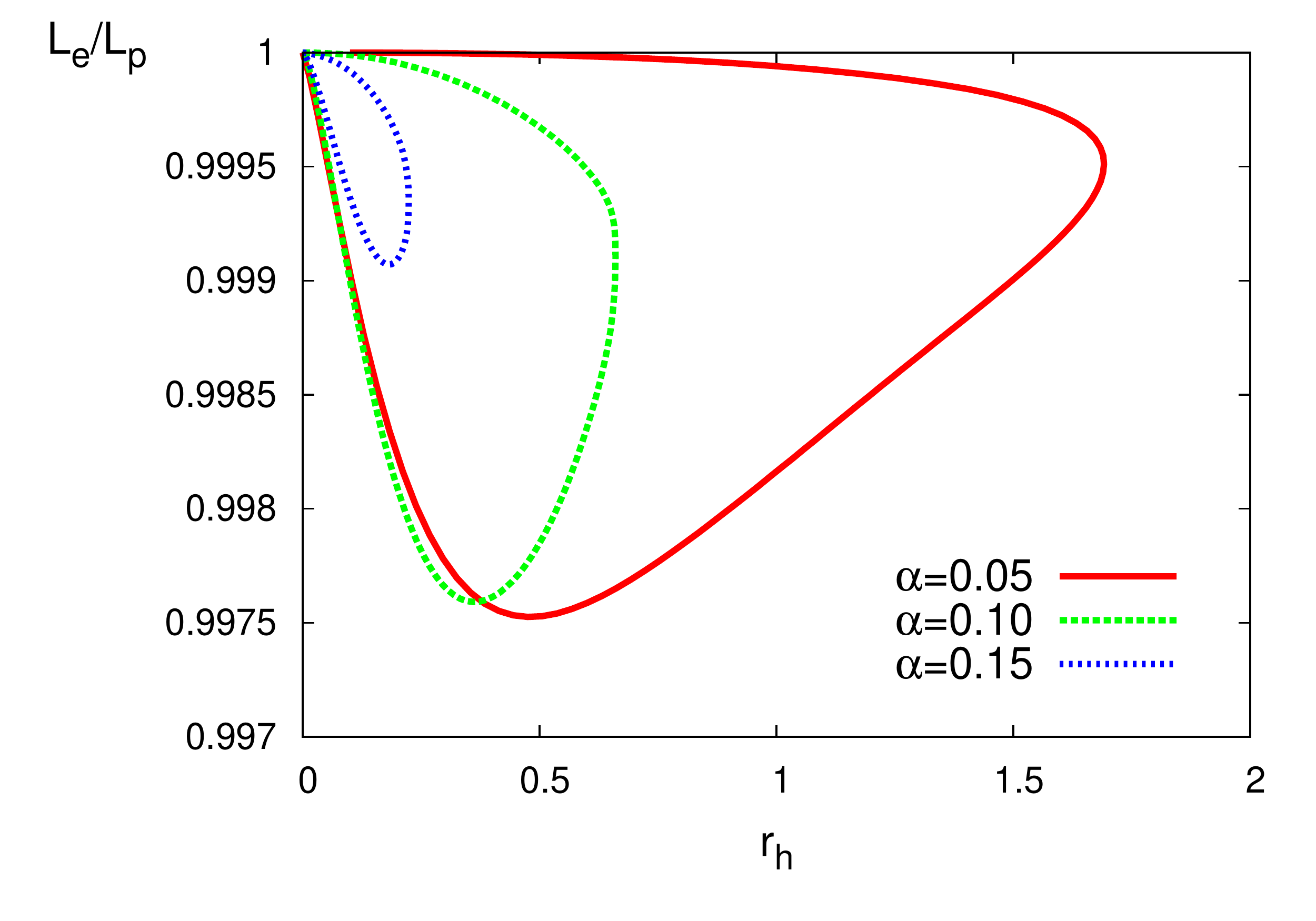}
\end{center}
\caption{\small
The value of the metric function $f_2$ on the event horizon
and the ratio $L_e/L_p$ (right plot) of the
$|n|=2$ S-A pair are shown as functions of the scaled event horizon
radius $r_h/\alpha$ for some set of values of  $\alpha$}.
\lbfig{fig5}
\end{figure}

The lower in energy branch of the solutions emerges
from the corresponding regular self-gravitating solution with a nonzero mass
as the horizon radius $r_h$ increases from zero.
This configuration can be viewed as a small Schwarzschild black hole,
immersed into the Skyrmion-anti-Skyrmion chains. In Fig.~\ref{fig1} we display the distribution of the
charge density $B_0$ \re{topcharge}, the third component of
the Skyrme field $\phi_3$ and the metric function $f$ for the $n=1$ single black hole with Skyrmion hair ($k=1$),
the S-A pair ($k=2$) and for the S-A-S chain ($k=3$) at $\alpha=0.15$.

Many features of the hairy black holes usually become more transparent if we introduce the
scaled ADM mass $M\alpha$ and scaled radial coordinate $r/\alpha$ \cite{Bizon:1992gb}.
The scaled mass of the configuration remains almost constant along the lower-in-mass  branch
while the Hawking temperature and the value of the metric function  $f_2(r_h)$ at the horizon
decreases, as the horizon radius $r_h$ increases, see Figs.~\ref{fig2}-\ref{fig5}.

An interesting feature shared by all BHs with Skyrmion-anti-Skyrmion hairs, is that they
always possess a prolate horizon, $L_e/L_p < 1$, although the horizon's deformation from
sphericity remains very small, as seen in Figs.~\ref{fig3},\ref{fig5}.  A tiny deformation is observed
for the  S-A-S chain with the winding $n=1$,
the strongest prolate deformation is seen
for the $n=1$ Skyrmion-anti-Skyrmion pair.
We observe that increasing of both the
gravitational coupling $\alpha$ and the winding $n$ reduces the deformation of the horizon.
Also, both the Skyrme field and the energy density do not vanish on the horizon, possessing an
angular dependence which is relatively large for most solutions we found.

Similar to the case of the spherically symmetric black holes
with Skyrme hair \cite{Luckock:1986tr,Droz:1991cx}, the lower branch always terminates at some critical
maximal value of the horizon radius $r_h^{(cr)}$. There it merges a secondary, upper (higher-mass) branch, which
extends backward in $r_h$, see Figs.~\ref{fig2}. The upper-branch solutions always
have higher entropy than the corresponding solutions on the lower-in-mass branch.
For the same value of the horizon radius, the deformation of the horizon is stronger on the
of the second branch, see  Figs.~\ref{fig3},\ref{fig5}.

As expected, the value of the critical horizon radius $r_h^{(cr)}$
decreases as the  gravitational coupling $\alpha$ grows, see Figs.~\ref{fig4},\ref{fig5}.
In the limit $r_h\to 0$ the solution of the upper branch approaches the corresponding
solution on the upper branch of the regular self-gravitating Skyrmion-anti-Skyrmion chains.
The black holes with Skyrmion hairs cease to exist at some critical maximal value of the gravitational
coupling $\alpha$.
The maximal value of the horizon radius $r_h^{(cr)}$ weakly depends on the winding number $n$, it sightly decreases
as $n$ increases.

\section{Conclusions}

We have considered new families of asymptotically flat
static hairy BHs in Einstein-Skyrme theory, which represent a black hole
immersed into the center of a chain of Skyrmions and anti-Skyrmions in interpolating order.
Analogous to their counterparts in the EYM theory \cite{Kleihaus:2007vf} and in the EYMH theory
\cite{Kleihaus:2000kv}, these solutions emerge from
the corresponding regular self-gravitating axially symmetric Skyrmion-anti-Skyrmion chains, as a small
event horizon radius $r_h$ is imposed via the boundary conditions.
Their domain of existence is restricted both by some maximal value of the $r_h$ and by a
maximal value of the
effective gravitational coupling $\alpha$.

In summary, concerning the dependence of the black holes with  Skyrmion-anti-Skyrmion hairs
on the gravity coupling constant and on the horizon raduis $r_h$, we
generally observe the picture that is very similar to the pattern
observed for the Einstein-Yang-Mills hairy black hole sphaleron solutions \cite{Kleihaus:2000kv,Ibadov:2005rb}.
The most important difference is that for
the black hole solution in the Einstein-Skyrme theory we never observe oblate deformations of the
horizon.

It would be interesting to investigate the stability of these new
black hole solutions with Skyrmion-anti-Skyrmion hairs. By analogy with the usual
stability analysis of the spherically symmetric black holes in the Einstein-Skyrme theory
\cite{Heusler:1991xx,Maeda:1993ap},
one can expect the existence of unstable  fluctuations on the upper branch.
However, the problem of systematic study of the spectrum of fluctuations of the fields
in the presense of the event horizon is a nontrivial technical task, which we leave for future study.

In the present work we have focused on the Skyrmion-anti-Skyrmion chains with two and three
constituents and with the winding number of each individual
component restricted to the lowest values $n=1,2$. As a direction for
future work, it would be interesting to study the higher charge
solutions, in particular the chains with charge 3 and 4 Skyrmions,
which do not possess axial symmetry \cite{Houghton:1997kg}.

he work here should be taken further by considering the isorotating generalization of the
Skyrmion-anti-Skyrmion chains, which will extend the study of the spinning black holes with Skyrme
hairs \cite{Herdeiro:2018daq}.
It would be also interesting to address the question of how the properties of the
static axially symmetric black hole solutions with chainlike Skyrmion-anti-Skyrmion hairs
will be modified in the asymptotically  AdS spacetime.
We hope to return elsewhere with a discussion of some of
these interesting problems.

\section*{Acknowledgements}
The work was supported by Ministry of Science and High Education of
Russian Federation, project FEWF-2020-0003. Computations were performed
on the cluster HybriLIT (Dubna).

 \begin{small}
 
 \end{small}

 \end{document}